# PARTICLES, WAVES AND VACUUM IN FIVE DIMENSIONS: A STATUS REPORT

Paul S. Wesson

1. Department of Physics and Astronomy, University of Waterloo, Waterloo, Ontario N2L 4G1, Canada.

2. Herzberg Institute of Astrophysics, National Research Council, Victoria, B.C. V9E 2E7, Canada

Abstract: Since the 5D canonical metric embeds all 4D vacuum solutions of Einstein's equations, I review its application to the cosmological 'constant', quantized particles, deBroglie waves, scalar fields and wave-particle duality. There are several ways to rationalize these things using an extra dimension. A possible explanation of wave-particle duality is that an observed particle manifests *two* isometries of flat 5D space in different 4D ways, one with waves and one without.

PARTICLES, WAVES AND VACUUM IN FIVE DIMENSIONS:

A STATUS REPORT

1.  Introduction

Attempts to model the properties of elementary particles using general relativity have a long history, and continue to the present [1 – 6]. While Einstein's theory was developed for physics on the large scale, it is the archetype of field theories with covariance, an attribute which should also apply on the small scale. There are tantalizing similarities between the two domains, such as the fact that the gyromagnetic ratio of a spinning black hole with electric charge is the same as that of the electron [5]. While nobody expects a classical field theory to reproduce all of the properties of a particle as described by quantum theory, the former may offer clues on how to resolve certain longstanding problems in the latter subject. These include the nature of antimatter [7, 8], how to superpose states corresponding to particles with different masses [9, 10], the electromagnetic radiation (or its lack) from accelerated charged particles, and other difficulties with offshell electrodynamics [11]. These are basic problems, and while quantum field theory works well in other regards, their resolution may need a theory with wider scope.

It is widely acknowledged that the unification of gravity with the interactions of particles is best approached via a theory with more dimensions than the standard four of spacetime. Five is the basic extension, and non-compactified versions of the Kaluza-Klein type, such as space-time-matter theory and membrane theory, agree with observations and are currently under close scrutiny [12]. Relevant 5D discoveries include: solitons which have unusual pressures that mimic the Poincaré stresses postulated for the stability of old particle models [13]; waves in the classical deSitter vacuum which are observable in ordinary 3D space [14]; exact 5D solutions which resemble the 4D charged black-hole ones of Reissner-Nordstrom but have a different combination of mass and charge [15]; and general expressions for the mass and electric and magnetic charges of 5D particles [16]. These are hints as to how an extra dimension can improve our understanding of particles.

Surprisingly, even the classical 4D vacuum shows new properties when it is embedded in 5D [12]. This is based on the *Theorem*: Any solution of the 4D Einstein field equations, without ordinary matter but with a finite energy density for the vacuum as measured by the cosmological 'constant', is also a solution of the 5D Ricci-flat field equations with a metric of the type known as canonical [17]. This metric is designated $C_5$, and should not be confused with the 5D Minkowski metric $M_5$. Some solutions of Einstein's 4D equations, such as the Kerr metric for a spinning black hole, can be embedded in $C_5$ but not in $M_5$. However, some solutions (with and without matter) can be embedded in both $C_5$ and $M_5$. An important example is the deSitter metric, in both its cosmological and local forms. The flat embedding was first achieved by Robertson [18] and applied to particles by Dirac [19]. It has many uses [20], and a new one will be examined below. Robertson also noted that the general cosmological metric named after him and Walker could also be embedded in $M_5$. This fact has been rediscovered by several workers up to the present [21], and means that even 4D standard cosmologies with matter can be embedded in a 5D space that is flat and empty. It should be noted, though, that $C_5$ has the advantage over $M_5$ of explicitly including the cosmological 'constant' $\Lambda$.

This parameter may actually be variable in 5D relativity, and in particular depend on the extra coordinate [12]. This effect is seen most clearly if a shift is applied to the extra coordinate in the straight canonical metric, when a hypersurface appears where the magnitude of $\Lambda$ diverges [22]. This hypersurface in space-time-matter theory is similar to that postulated in membrane theory, and the two approaches actually have a similar mathematical structure [23]. In $\Lambda$-dominated $C_5$ cosmology with $\Lambda > 0$, the early universe is necessarily inflationary [24, 25]. The $\Lambda$-divergent hypersurface of $C_5$ with $\Lambda < 0$ provides a kind of 'groove' in the manifold that can be applied to particle physics [26]. In short, a shift in the extra coordinate of the canonical metric introduces structure to the vacuum.

In what follows, I wish to ask the question: What aspects of particle physics can be better understood using 5 dimensions rather than 4? Some of the problems with conventional theory were mentioned above, and it is reasonable to look for their resolution in an extra dimension. Following recent results in the literature, it will be assumed that in general the scalar field associated with the extra dimension is responsible for particle mass, that in the special case of a constant scalar potential the metric has the canonical form, and that all particles (massive as well as massless) follow null paths in 5D. The notation is standard, with upper-case Latin letters running 0, 123, 4 for time, space and the extra dimension, while lower-case Greek letters run 0, 123. The speed of light $c$, Newton's gravitational constant $G$ and Planck's constant of action $h$ are all set to unity, except where they are made explicit to aid physical understanding.

2. <u>Particles, Waves and Vacuum in Five Dimensions</u>

The most baffling property of matter on the small scale is that it exists as localized particles and extended waves. Wave-particle duality indicates that there is something special about the underlying medium, the vacuum. The latter, like its predecessor the aether, appears to possess puzzling properties. On the small scale, space appears to be filled with intense fields attendant on normal particles, and is also full of creating and annihilating virtual particles, though the sum of these things produces negligible curvature in spacetime. On the large scale, space is full of massive galaxies, but their average density is only a fraction of the energy density as measured by the cosmological constant, and the combination of ordinary matter and vacuum produces a measurable curvature in spacetime.

Other problems will be examined below, but we begin with asking: How can the vacuum support phenomena that are both particle-like and wave-like?

The canonical metric, as mentioned above, embeds the 4D vacuum solutions of general relativity in 5D. The extra coordinate $x^4 = l$ can be either spacelike or timelike. The scalar field $\Phi$, which is associated with $l$ and is believed to be connected to the generation of rest mass, can be flattened by an appropriate choice of coordinates so the extra metric coefficient becomes $g_{44} = \pm 1$. Similarly, the electromagnetic field is removed by setting $g_{4\alpha} = 0$. And the gravitational field is factorized by a quadratic in $x^4 = l$, leaving the rest dependant only on the spacetime coordinates $g_{\alpha\beta}(x^\gamma)$. The result is a simple 5D metric, which by substitution into the Ricci-flat 5D field equations is found to also satisfy the 4D Einstein equations, with a cosmological constant $\Lambda$ that depends on the curvature scale $L$ of spacetime. Thus:

$$dS^2 = (l/L)^2 ds^2 \pm dl^2 \qquad (1.1)$$

$$ds^2 = g_{\alpha\beta}(x^\gamma) dx^\alpha dx^\beta \qquad (1.2)$$

$$\Lambda = \mp 3/L \quad . \qquad (1.3)$$

Here the sign of $\Lambda$ is correlated with the sign of the extra dimension.

This choice of signs affects the 4D physics. To illustrate, consider the embedding in (1) of the local deSitter solution, whose 4D metric is given by

$$ds^2 = (1-\Lambda r^2/3)dt^2 - (1-\Lambda r^2/3)^{-1} dr^2 - r^2(d\theta^2 + \sin^2\theta d\phi^2) \quad . \qquad (2)$$

It has been known for a long time that that this solution can be embedded in 5D Minkowski space $M_5$ as well as in the canonical space $C_5$ of (1). In flat 5D, (2) resembles a pseudosphere of radius $L$. The Gaussian curvature is $K = -1/L^2$ for $\Lambda > 0$ and $K = +1/L^2$ for $\Lambda < 0$. The more physical Ricci or curvature scalar is related by $K = -R/12$. The Ricci scalar for a perfect fluid measures the contending influences of the density $\rho$ and pressure $p$ via the combination $(\rho - 3p)$, which is zero for photons or ultrarelativistic matter. The cosmological constant in general relativity can be connected to the properties of a vacuum fluid with the equation of state $\rho_v = \Lambda/8\pi = -p_v$ which has $R = 4\Lambda$.

The universe on the macroscopic scale appears to have $\Lambda > 0$ (though $\Lambda < 0$ is possible on microscopic scales), where the corresponding radius of curvature is of order $10^{28}$ cm. A cosmological version of (2) with $\Lambda > 0$ can be obtained by a coordinate transformation, and reads

$$ds^2 = dt^2 - \exp\left[2(\Lambda/3)^{1/2} t\right]\left[dr^2 + r^2\left(d\theta^2 + \sin^2\theta d\phi^2\right)\right] \quad . \tag{3}$$

This deSitter model is characteristic of inflationary cosmology, and we will consider its 5D complex generalization below as a model for deBroglie waves. Here we note that while (2) and (3) can be embedded in $M_5$, other vacuum spacetimes such of that of Schwarzschild cannot be embedded in flat manifolds of less than 6 dimensions. But since *all* 4D vacuum metrics can be embedded in $C_5$ of (1), canonical space is the natural basis for modelling particles surrounded by vacuum and waves moving through vacuum.

Both types of behaviour follow from null-paths in (1) specified by $dS^2 = 0$. This 5D condition describes the paths of all particles in 4D, even massive ones, and replaces the conventional definition of causality specified by $ds^2 \geq 0$ and based on photons. The 5D null-path in (1) gives two types of motion:

$$l = l_* e^{\pm s/L} \quad , \quad l = l_* e^{\pm is/L} \quad . \tag{4}$$

Here $l_*$ is an arbitrary constant, and both types of motion are reversible in the extra dimension. The monotonic or particle-like motion, and the wave-like motion, arise respectively when the extra dimension is spacelike or timelike.

There are actually 4 possible modes in the $C_5$ metric rather than 2. The first part of (1.1) is $(l/L)^2 ds^2$, and is invariant under $L \to iL, s \to is$. By (4), when $x^4 = l$ describes wave motion, the form of $l$ is the same as that of the wave function in old wave mechanics, provided $L = h/mc$ is the Compton wavelength of the associated particle with mass *m*. It may actually be shown that the extra component of the 5D geodesic equation, which is the equation of motion for $l(s)$, is equivalent to the Klein-Gordon equation of quantum mechanics [26]. We will return to the Klein-Gordon equation later, and also discuss deBroglie or matter waves, which are peculiar in that they involve velocities which exceed lightspeed. In special relativity, velocities exceeding *c* imply imaginary values for *m*, properties which are usually considered quaint curiosities. However, such a tachyonic sector is actually present via the aforementioned $L \to iL$, $s \to is$ in $C_5$. Due to the form of the metric, the imaginary mode and the real mode are degenerate in $C_5$. This is different from the situation in $M_5$, though it should be noted that any 5D metric with signature $(+---+)$ and a null interval admits, in a formal sense, velocities in 3D which exceed lightspeed.

There is another version of the canonical metric (1) which is obtained by shifting the extra coordinate along its own axis. We note this here, though its application will be deferred to below. Likewise, for a discussion of the field equations and their consequences. It can be mentioned here, however, that the 5D motions (4) affect what is observed in 4D, because the orbits $l = l(s)$ perturb the surface of spacetime. This is most clearly seen in the case of (2) with $\Lambda < 0$, where spacetime can be visualized as the surface of a sphere which is rippled with the waves of (4). The waves are described by the coordinate $x^4 = l$ rather than its associated potential $g_{44} = \pm\Phi^2$ because the latter has been absorbed by the choice of coordinates. Our 4D spacetime is technically the surface $l = L$ in the 5D manifold (1.1). For $\Lambda > 0$, particles wander away from or towards a given *l*-surface at a rate governed by the magnitude of $\Lambda$, and it may be shown that while their masses may vary in accordance with the law of conservation of momentum, the motion appears geodesic in the conventional sense in the appropriate coordinates [12, 17]. For $\Lambda < 0$, waves oscillate around a given *l*-surface with an arbitrary amplitude but a wavelength governed by the magnitude of $\Lambda$, and accordingly the 4D dynamics will in general have a certain fuzziness. These characteristics are maintained when the metric (1) is shifted by a constant, $l \rightarrow (l - l_0)$. But there is an important change in the nature of $\Lambda$ as determined by Einstein's equations in the surface of spacetime [22]. Thus:

$$dS^2 = \left[(l - l_0)/L\right]^2 ds^2 \pm dl^2 \qquad (5.1)$$

$$ds^2 = g_{\alpha\beta}(x^\gamma) dx^\alpha dx^\beta \qquad (5.2)$$

$$\Lambda = \mp \frac{3}{L^2}\left(\frac{l}{l - l_0}\right)^2 \qquad . \qquad (5.3)$$

Clearly a hypersurface $(l = l_0)$ now exists where the magnitude of $\Lambda$ is divergent. This hypersurface, while similar to the singular surface of membrane theory, turns out to have different properties [12]. Notably, the periodic $C_5$ wave can be trapped in the 'groove' formed by the negative divergence of $\Lambda$, but can traverse this hypersurface. This behaviour suggests quantization. Accordingly, we turn our attention to the possibility of modelling a particle not as a point but as a tiny ball of trapped waves.

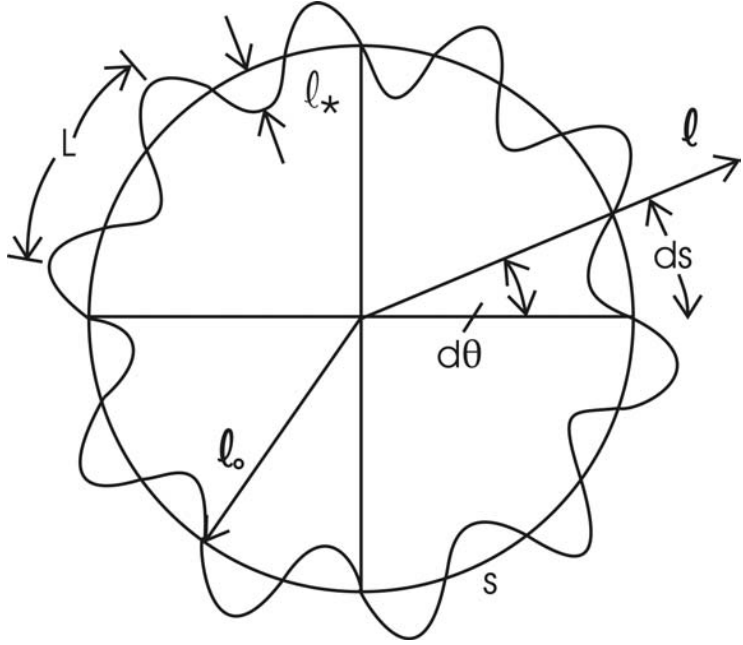

Figure 1: A vacuum wave with wavelength $L$ and amplitude $l_*$ whose locus is a circle with radius $l_0$ in the fifth dimension. The fifth coordinate $l$ and the angle $\theta$ are used to describe the closed circumference $s$ of spacetime.

To model a quantized particle, consider the shifted-canonical metric (5) with a timelike extra dimension $(\Lambda < 0)$ and a path which is 5D null $(dS^2 = 0)$. The orbit in the $l/s$ plane is given by

$$l = l_0 + l_* e^{\pm is/L} \quad , \tag{6}$$

where $l_0$ is the shift, $l_*$ is the amplitude and $L$ is the wavelength of the wave. The surface of spacetime $s$ is closed and the topology can be taken as circular (see Figure 1). If the wave is associated with a particle of mass $m$, $L$ is the Compton wavelength $L = h/mc$. As noted above, the negative divergence of $\Lambda$ at $l = l_0$ effectively traps the wave there. Suppose there are *n* wavelengths in the circumference, so $nL = 2\pi l_0$, which combined with the previously-noted Compton wavelength gives

$$l_0 = n(\hbar/mc) \quad . \tag{7}$$

The action as usually defined by an integral over *s* can be rewritten as an integral over the angle $\theta$ where $ds = l_0 d\theta$, and then this can be simplified by eliminating $l_0$ using (7). That is,

$$I = \int mc\,ds = \int mcl_0 d\theta = \int n\hbar d\theta \quad . \tag{8}$$

The result is the standard quantization rule:

$$I = \int_0^{2\pi} n\hbar d\theta = nh \quad . \tag{9}$$

This result is the same as that in a previous model of similar type [26], but now the interpretation of the wave and the quantum number *n* are somewhat different (a standing wave and a running wave give the same answer). The *l*-wave (6) which leads to the quantization rule (9) exists because of the properties of the vacuum, which has positive pressure and negative energy density for $\Lambda < 0$.

It is also possible, provided $\Lambda < 0$, to have waves which propagate freely in 3D rather than being trapped in 4D. The latter type of wave arises essentially from an embedding in 5D of the local version of the deSitter solution in general relativity. However, the cosmological deSitter solution (3) is also useful, and has been applied in several different contexts [12, 14, 18, 20, 25]. One reason is that it can be embedded in both $M_5$ and $C_5$, as noted before. Actually, depending on the choice of coordinates, the 5D deSitter solution exists in several different versions, or isometries. An instructive form $C_5$ is:

$$dS^2 = \frac{l^2}{L^2}\left\{dt^2 - \exp\left[\pm\frac{2i}{L}(t+\alpha x)\right]dx^2 - \exp\left[\pm\frac{2i}{L}(t+\beta y)\right]dy^2\right.$$

$$\left. - \exp\left[\pm\frac{2i}{L}(t+\gamma z)\right]dz^2\right\} + dl^2 \quad . \tag{10}$$

This describes a wave propagating through ordinary 3D space. The dimensionless constants $\alpha, \beta, \gamma$ are akin to wave numbers and are arbitrary. But the frequency is fixed by the solution (which satisfies $R_{AB} = 0$), and depends on the constant length $L$. This is related to the cosmological constant $\Lambda$, which for (10) is $\Lambda = -3/l^2$ in general or $\Lambda = -3/L^2$ on the hypersurface $l = L$. Restoring conventional units for the speed of light $c$, the phase velocity of the wave along (say) the x-axis is $c/\alpha$ and can in principle take on any value.

Superluminal velocities occur in quantum mechanics primarily in connection with deBroglie waves and the Klein-Gordon equation, which is the relativistic version of the Schrodinger equation. Much has been written about these topics, some of it verging on the mystical. This because certain aspects of the theory appear to lead to counter-intuitive results. An example is the collapse of the wave function when a particle interacts with an observer, leading to a violation of unitarity and an apparent loss of information. DeBroglie waves follow automatically when the expressions for the energy of a particle $(E = mc^2)$ and a wave $(E = hf)$ are combined. A consequence of the combined wave/particle dynamics is that two velocities appear, one for each entity, which are related by $vw = c^2$ [20]. Obviously, if one of $v, w$ is less than $c$ then the other must exceed $c$. This can be understood as a consequence of vacuum dynamics in a canonical-type metric in 5D, as shown above. But it is also, in a way, consistent with the Minkowski metric in 4D, by the following argument. Consider two frames, one of which moves with respect to the other at velocity $v$. The Lorentz transformations give a relation between the time measured in one frame and the time and location in the other, namely $t' = (t - vx/c^2)(1 - v^2/c^2)^{-1/2}$. Suppose a wave is emitted in the one frame at $t' = 0$. Then it is observed in the other frame at $t = vx/c^2$. This is location-dependent, and means that the wave propagates through the latter frame at a speed which is formally $w = x/t = c^2/v$. Therefore $vw = c^2$, as noted. This argument, applied to a particle and a wave, implies that the wave appears to have a superluminal velocity. However, while this phenomenon may be *consistent* with $M_4$, it is better *explained* by the $C_5$ metric (10).

The Klein-Gordon equation may likewise be derived in a better way than usual. In the standard approach, the 4-velocities $u^\alpha \equiv dx^\alpha/ds$ are normalized via $u^\alpha u_\alpha = 1$, and this is multiplied throughout by an arbitrary constant $m^2$ to give the normalization condition $p^\alpha p_\alpha = m^2$ for the momenta. These are then replaced by operators, acting on a hypothetical complex function $\psi$ known as the wave function:

$$p_\alpha = (h/i\psi)\partial\psi/\partial x^\alpha \quad . \tag{11}$$

With this prescription, the normalization condition $p^\alpha p_\alpha = m^2$ gives

$$\Box\psi + (c/h)^2 m^2 \psi = 0 \quad . \tag{12}$$

This is the Klein-Gordon equation ($\Box\psi \equiv g^{\alpha\beta}\psi_{,\alpha;\beta}$ where a comma denotes the partial derivative and a semicolon denotes the covariant derivative). It is an operator-based rep-resentation of the 4-velocities, where the particle mass $m$ is introduced from outside and the complex nature of $\psi$ reflects the mixed signature of the metric $(+---)$. A more transparent derivation is as follows. Form a dimensionless action by measuring the interval in units of the Compton wavelength, and use this to define a wave function:

$$I \equiv \int (h/mc)^{-1} ds \quad , \quad \psi \equiv e^{iI} \quad . \tag{13}$$

The first derivative of $\psi$ gives (11) above. The second derivative, taken covariantly if the spacetime is curved, splits into a real part and an imaginary part. One of these parts gives $p^{\beta}_{;\beta} = 0$, the standard conservation law for the momenta (equivalent to the geodesic equation of motion if $m$ is constant). The other part gives (12) above, the Klein-Gordon equation. An advantage of this method is that it reveals the possibility, at least in principle, of a kind of 'mirror' situation in which $m \to im$ in the previous analysis. That is, the Klein-Gordon equation now involves a real $\psi$ function and is satisfied by a particle with $m^2 < 0$. Matter consisting of particles with $m^2 < 0$ should be distinguished from conventional antimatter, which falls downward in a gravitational field like ordinary matter. It should also be distinguished from matter consisting of particles with $m < 0$, as admitted by conformally-invariant 4D gravity [27]. Particles with $m^2 < 0$ are instead allowed by 5D gravity, where the conformal structure of the metric implies a proportionality $\Lambda \sim m^2$ and $\Lambda$ changes sign when the extra dimension changes sign [28]. The effects of changing the nature of the fifth dimension in canonical-type 5D metrics were mentioned near the beginning of this section. However, it is instructive to make some brief comments for metrics of more general type.

In 5D metrics, the sign of the last term is often inserted 'by hand', but it can also change 'naturally'. The last component of the metric tensor $g_{44}$ represents the scalar field typical of 5D relativity, but not much is known about its physics, and the sign of $g_{44}$ is frequently determined by working through the algebra to obtain a solution of the field equations $R_{AB} = 0$. Solutions are known where $g_{44}$ has one sign or the other [12]; and for some solutions, like the solitons, it can have either sign [29]. It is possible to discern three ways in which $g_{44}$ can change sign: (i) as a result of a symmetry in the field equations, notably involving the $R_{44}$ component; (ii) as a consequence of passing through an horizon, perhaps due to the dynamics; (iii) as the effect of a complex scalar field changing from real to imaginary, connected with a switch between monotonic and oscillatory modes. In the present work, the focus is on particles, waves and the vacuum. Even in the $C_5$ class of metrics, where the scalar field is set to a constant, a change in the sign of $g_{44}$ is connected directly to a change in the sign of the cosmological 'constant'. It is of critical importance to understand in more detail the connection between $g_{44}$ and $\Lambda$.

The connection between the scalar field and the vacuum can be understood by considering the more general metric

$$dS^2 = \frac{l^2}{L^2} g_{\alpha\beta}\left(x^\gamma, l\right) dx^\alpha dx^\beta + \varepsilon \Phi^2 dl^2 \quad . \tag{14}$$

Here $\varepsilon = \pm 1$ fixes the timelike (+) or spacelike (-) nature of the extra dimension. There are known many solutions for metric (14) of the 5D field equations $R_{AB} = 0$. These 15 relations split into 3 sets: one set is a form of Einstein's equations $G_{\alpha\beta} = 8\pi T_{\alpha\beta}$, where the energy-momentum tensor $T_{\alpha\beta}(\partial g_{\alpha\beta}/\partial l)$ is induced by the extra dimension and determines the density $\rho$ and pressure $p$ of matter; the second set comprise 4 conservation laws $P^{\beta}_{\alpha;\beta} = 0$, which determine the behaviour of matter currents in spacetime; the last component of the field equations is a wave equation $\Box\Phi = Q(\partial g_{\alpha\beta}/\partial l)$, which determines the behaviour of the scalar potential in terms of a source that depends on the matter.

An especially informative class of solutions for (13) is of cosmological type, with flat 3D sections. The metric coefficients are

$$g_{00} = +1, \quad g_{11} = g_{22} = g_{33} = -t^{2/\alpha} l^{2\alpha/(1+\alpha)}, \quad g_{44} = -\Phi^2 = \alpha^2(1-\alpha)^{-2} t^2 \quad . \quad (15)$$

These solutions describe Friedmann-Robertson-Walker models on the hypersurfaces $l =$ constants. The class was found by Ponce de Leon [30] and has been studied by Wesson and others, who have evaluated its physical properties [12]. The density and pressure are given by $8\pi\rho = 3/\alpha^2 T^2$ and $8\pi p = (2\alpha - 3)/\alpha^2 T^2$, where $T \equiv tl$ is cosmic time and $\alpha$ is a dimensionless constant ($\alpha = 3/2$ and $\alpha = 2$ describe dust-like and radiation-like matter). However, while these quantities embody the usual physical properties of matter, the currents which appear in the field equations are somewhat different in form. Some tedious algebra gives

$$P^0_0 = -3/\alpha T, \qquad P^1_1 = P^2_2 = P^3_3 = (\alpha - 3)/\alpha T \quad . \quad (16)$$

In terms of these, the physical density and pressure are

$$8\pi\rho = (P^0_0)^2/3, \qquad 8\pi p = P^0_0(P^0_0 - 2P^1_1)/3 \quad . \quad (17)$$

These properties of matter also figure in the last component of the field equations $(R_{44} = 0)$, which has no analog in general relativity and deserves attention.

Evaluating $R_{44} = 0$ using previous relations shows that the scalar field obeys

$$\Box \Phi = (3/2)(8\pi)(\rho + p)\Phi \quad . \tag{18}$$

This reveals that the evolution of the scalar field depends on what is sometimes called the inertial mass density $(\rho + p)$, a name which follows from the fact that in FRW models the rate of change of the density is proportional to this combination, which thereby determines the stability of matter. It should be distinguished from the gravitational mass density $(\rho + 3p)$, a name which follows from the fact that the acceleration of a test particle is proportional to the noted combination, which thereby determines the attractive nature of gravity. The two types of density obey simple relations in the case of FRW models, and it is instructive to recall them here. They can be derived by manipulating the Friedmann equations, which are Einstein's field equations for a homogeneous and isotropic fluid that is expanding with scale-factor $R(t)$. Using an overdot to denote the time derivative, the two relations concerned are:

$$\dot{\rho} = -(\rho + p)\left(\frac{3\dot{R}}{R}\right) \tag{19}$$

$$\ddot{R} = \frac{-(4/3)\pi R^3 (\rho + 3p)}{R^2} \quad . \tag{20}$$

Comparing these to (18) above, it is apparent that the $\Phi$-field is a non-gravitational type of interaction.

The results derived above can be summarized in a couple of comments which have wider implications. First, the common properties of matter $(\rho, p)$ depend on more basic currents $(P_0^0, P_1^1)$, a situation reminiscent of particle physics, where the observed particles are often mixtures of more fundamental states. Second, the scalar field obeys a wave equation which resembles the Klein-Gordon equation of quantum theory more than anything else, and shows that the field becomes 'free' in the limit $(\rho + p) \to 0$, the vacuum state.

3. <u>Summary and Discussion</u>

The material we have covered in the preceding section is diverse, but is all based on the 5D canonical metric. In its standard form (1) it provides an embedding for 4D vacuum solutions, including the deSitter metric in its local form (2) and cosmological form (3). The cosmological 'constant' $\Lambda$ is necessarily finite, and the topology of spacetime is either open ($\Lambda > 0$) or closed ($\Lambda < 0$). 5D null-paths cause the extra coordinate $x^4 = l$ to vary with 4D proper time, in either a monotonic ($\Lambda > 0$) or oscillatory ($\Lambda < 0$) manner (4). When a shift is included in the $C_5$ metric (5), $\Lambda$ is no longer constant but becomes a function of $l$, which via $l = l(s)$ means that $\Lambda$ is variable in spacetime. For $\Lambda < 0$, the wave defined by $l(s)$ of (6) is trapped in the 'groove' formed by the negative divergence of $\Lambda$. The closed spacetime behaves like a particle, with the standard quantization rule (9). An alternative form (10) of $C_5$ with $\Lambda < 0$ displays a running wave in ordinary 3D space, whose phase velocity may be arbitrarily large, as with deBroglie waves. The Klein-Gordon equation (12) of 4D wave mechanics can be derived in several ways, but is equivalent to the extra component of the 5D geodesic equation for metrics of $C_5$ type. For metrics of more general type (14) there is a class of solutions (15) whose matter currents (16) are defined by the 5D field equations, but yield the ordinary 4D density and pressure (17). For these solutions, the scalar field obeys an equation (18) similar to that of Klein-Gordon. However, its source is not the one typical of gravity (20), but the one that characterises the inertial properties of matter.

The results summarized above are like the pieces of a puzzle, and it should be admitted that they still need to be fused into a coherent whole. The astute reader will no doubt see where further work is required towards this end.

The biggest problem for 5D physics concerns the twin issues of the magnitude of the cosmological 'constant' and wave-particle duality. These issues overlap insofar as the canonical metric is concerned. It is widely believed that elementary particles involve intense vacuum fields, and that the real ones we see interact with an energetic ocean of virtual ones we do not see. This implies large values of $\Lambda$ on small scales of order $10^{-12}$ cm. Yet when particles move around in 3D over laboratory distances, they do not show significant effects of the severe curvature that would follow from an energetic vacuum. And cosmological observations imply a small value for the magnitude of $\Lambda$ on large scales of order $10^{28}$ cm. In addition to this puzzle, it has been known for ages that particles sometimes show classical, point-like behaviour and sometimes quantum, wave-like behaviour. The addition of a fifth dimension can in principle resolve both of these issues, because $\Lambda$ as a measure of the size of the local potential can vary from scale to scale, and $l = l(s)$ for the motion has both monotonic and oscillatory modes depending on the sign of $\Lambda$. However, it is difficult in practise to find a single prescription which fits all the facts.

Canonical space can be used, as we have seen, to model various aspects of a particle and its motion. The particle as an object can be modelled as a tiny, quantized ball of vacuum energy with $\Lambda < 0$. The extra coordinate modulates the 4D part of the metric, and evolves smoothly if $\Lambda > 0$, but is oscillatory and formally the same as the conventional wave function if $\Lambda < 0$. One possibility, which matches observations, is this: Suppose *both* modes are realized, in such a way as to exhibit both particle and wave behaviour, and that the positive and negative values of $\Lambda$ *cancel* to leave almost flat spacetime. This, admittedly, sounds strange. But no more so than some other interpretations of quantum mechanics. If this picture should turn out to be correct, and the combined metric is isometric to the 5D Minkowski one, it would mean that a "wavicle" is two simultaneous realizations of flat space, one with waves and one without.

Acknowledgements

This work grew out of an earlier collaboration with A. P. Billyard. Thanks for comments go to various members of the Space-Time-Matter group (http://astro.uwaterloo.ca/~wesson).